\documentclass[10pt]{article}

\usepackage{amsmath}
\usepackage{amssymb}

\usepackage{graphicx}

\usepackage{cite}
\usepackage{breakcites}

\usepackage{color}

\usepackage[labelfont=bf,labelsep=period,justification=raggedright]{caption}

\date{}

\begin{document}
\bibliographystyle{plain} 

\begin{flushleft}
{\Large
\textbf{Revealing missing parts of the interactome}
}
\\
Ryan W. Solava and
Tijana Milenkovi\'{c}$^{\ast}$
\\
 Department of Computer
  Science and Engineering, ECK Institute for Global Health, and Interdisciplinary Center for Network
  Science and Applications \\
  University of Notre Dame, Notre Dame, IN 46556, USA
  
$^{\ast}$E-mail: tmilenko@nd.edu
\end{flushleft}

\section*{Abstract}

Protein interaction networks (PINs) are often used to ``learn'' new
biological function from their topology. Since current PINs are noisy,
their computational de-noising via link prediction (LP) could improve
the learning accuracy. LP uses the existing PIN topology to predict
missing and spurious links.  Many of existing LP methods rely on
shared \emph{immediate} neighborhoods of the nodes to be linked. As
such, they have limitations. Thus, in order to \emph{comprehensively}
study what are the topological properties of nodes in PINs that
dictate whether the nodes should be linked, we had to introduce novel
\emph{sensitive} LP measures that overcome the limitations of the existing
methods.

We systematically evaluate the new and existing LP measures by
introducing ``synthetic'' noise to PINs and measuring how well the
different measures reconstruct the original PINs. Our main findings
are: 1) LP measures that favor nodes which are \emph{both}
``topologically similar'' \emph{and} have large shared \emph{extended}
neighborhoods are superior; 2) using more network topology often
though not always improves LP accuracy; and 3) our new LP measures are
superior to the existing measures. After evaluating the different
methods, we use them to de-noise PINs. Importantly, we manage to
improve biological correctness of the PINs by de-noising them, with
respect to ``enrichment'' of the predicted interactions in Gene
Ontology terms.  Furthermore, we validate a statistically significant
portion of the predicted interactions in independent, external PIN
data sources.

Software executables are freely available upon request.

\section*{Introduction}

\subsection*{Motivation and background}

Networks (or graphs) model real-world phenomena in many domains.  We
focus on biological networks, protein-protein interaction (PPI)
networks in particular, with the goal of identifying missing and
spurious links in current noisy PPI networks.  Nonetheless, our study
is applicable to other network types as well.  In PPI networks, nodes
are proteins and two nodes are connected by an edge if the
corresponding proteins interact in the cell.  We focus on these
networks, since it is the proteins (gene products) that carry out the
majority of cellular processes and they do so by interacting with
other proteins.  And this is exactly what PPI networks model.

High-throughput methods for PPI detection, e.g., yeast two-hybrid
(Y2H) assays or affinity purification followed by mass spectrometry
(AP/MS), have produced PPI data for many species
\cite{GiotSci03, Stelzl05, Yu2008, Simonis2009, BIOGRID}.  However,
current networks are \emph{noisy}, with many missing and spurious
PPIs, due to limitations of biotechnologies
as well as human biases 
\cite{Mering02, Stumpf05, Vidal05, Collins07, Wodak09}.
AP/MS is estimated to have a 15-50\% false positive rate and a 63-77\%
false negative rate \cite{Edwards02}.  Similar holds for Y2H, though
PPIs obtained by Y2H are still more precise than literature-curated
PPIs supported by a single publication \cite{Venkatesan2009}.

Analogous to genomic sequence research, biological network research is
promising to revolutionarize our biological understanding: prediction
of protein function and the role of proteins in disease from PPI
network topology has already received much attention
\cite{Sharan2006, Sharan2007, Sharan2008, Barabasi_Oltvai04, Sharan10}.
However, the noisiness of the network data is an obstacle on this
promising avenue, as it could lead to incorrect predictions.
Computational de-noising of current PPI network data by identifying
missing and spurious links could improve the quality of topology-based
predictions and consequently save resources needed for experimental
validation of the predictions.  Hence, we aim to test how well we can
decrease the noise in PPI data via link prediction (LP).

LP typically uses the existing topology of the network to predict
missing and spurious links
\cite{liben2003link, Getoor2005, LibenNowell2007, Lichtenwalter2010, Lu2010, Rattigan2005, Sarac2012}.
Alternativelly, one network type, e.g., functional interactions, can
be used to predict another network type, e.g., physical PPIs
\cite{Sarac2012}.  LP consists of unsupervised or supervised
approaches that use some measure of the topology of the nodes to be
linked \cite{Lichtenwalter2010}. For example, it may be desirable to
link nodes with high degrees as measured by preferential attachment
\cite{Newman2001}, nodes that share many neighbors as measured by
Jaccard \cite{Salton1984} or Adamic/Adar coefficients
\cite{Adamic2003}, or ``important'' nodes that interact with many
other ``important'' nodes as measured by PageRank \cite{Brin1998}.
Both supervised and unsupervised LP methods have their
(dis)advantages.  Though supervised methods can outperform
unsupervised ones, much of previous research has focused on
unsupervised LP, since many factors that might influence supervised LP
have not been well understood
\cite{Lichtenwalter2010, liben2003link,Lu2010}.

There are some limitations to the existing LP measures.  While many
capture only the topological information contained in the
\emph{immediate} network neighborhood of nodes to be linked,
significant amount of the information is available in the rest of the
network that could improve LP accuracy.  Hence, more sensitive
measures that capture deeper network topology are needed. We recently
generalized the idea of shared immediate neighborhoods to shared
\emph{extended} neighborhoods in the context of network clustering and
showed that including more network topology resulted in biologically
superior clusters \cite{Solava2012}.  So, it is reasonable to test
whether including more topology will be effective for LP as well.

Also, existing shared neighborhood-based measures can predict 
a link only between
nodes that are within the shortest path \emph{distance of two} from
each other, whereas it might be beneficial to link nodes which are
\emph{more distant}.
Preferential attachment-based measures can achieve this, but they
again capture only the immediate neighborhoods of the nodes to be
linked.  A shortest path-based LP method exists which can also connect
distant nodes in the network but which can at the same time capture
deeper network topology. However, this method is computationally
expensive \cite{Kuchaiev2010}.  Hence, we introduce a sensitive
measure of the topological similarity of \emph{extended} neighborhoods
of two nodes that addresses all of the above issues, and we use it
with the hypothesis that nodes that are topologically similar should
be linked together.

Another drawback of the existing methods is as follows.  It might be
more efficient to predict the existence of a link between two nodes by
explicitly measuring the topological position of an \emph{edge} (or
equivalently a non-edge) rather than by measuring the position of each
of the two \emph{nodes} individually, as the current methods do
\cite{Lichtenwalter2012}.  Thus, we propose a new, sensitive measure
of the network position of an \emph{edge} and a \emph{non-edge}, which
counts the number of subgraphs that the two nodes in question
participate in \emph{simultaneously}, and we use it with the
hypothesis that nodes that participate in many subgraphs and thus have
large and dense \emph{extended} shared neighborhoods should be linked
together.

\subsection*{Our approach}

We study several PPI networks of yeast, the best studied species to
date, obtained by different experimental methods for PPI detection,
and we apply our new as well as commonly used existing LP measures to
the networks to de-noise them.  Given a network, we aim to study the
topologies of each node pair in the network with respect to the given
LP measure, in order to determine which of the node pairs should be
connected. We compare the different LP methods in systematic
receiver-operator curve and precision-recall settings.  For each of
the two method comparison settings, we perform two types of evaluation
tests.  First, we introduce synthetic noise in the given PPI network
by randomly removing a percentage of edges from the network, with the
goal of measuring how well the given method can reconstruct the
original network, using the original PPIs as the ground truth data.
Second, given the availability of low-confidence PPI data for one of
the studied networks, we apply the given method to this network and
use the corresponding low-confidence PPI data as the ground truth data
when evaluating the method.

We study the effects on LP accuracy of the ``topological similarity''
as well as size of the shared \emph{extended} neighborhoods of nodes,
where the nodes \emph{can} be distant in the network.  Also, we study
what amount of network topology should be used for LP.  We find that
LP measures which favor nodes which are \emph{both} topologically
similar and which have large shared extended neighborhoods are
superior. We show that using more network topology often though not
always increases LP accuracy.  Importantly, we show that our new LP
measures are statistically significantly superior to the existing
ones.  Alarmingly, we find that receiver-operator curve and
precision-recall method comparison frameworks do not necessarily
agree, which has important implications for the LP community.

After we evaluate the different methods, we apply the methods to the
PPI networks to de-noise them, and we evaluate the quality of the
de-noised networks in two ways. First, we compute their biological
correctness by measuring the ``enrichment'' of predicted edges in Gene
Ontology (GO) terms \cite{Go00}. Importantly, we show that our new LP
measures as well as some of the existing measures improve the
biological correctness of the PPI networks by de-noising them. Second,
we search for the predicted interactions in an external, independent
PPI data source, and in this way, we validate a significantly large
portion of the predictions, further confirming the biological
correctness of the de-noised networks.

\section*{Methods}\label{sect:methods}

We study multiple \emph{S.  cerevisiae} PPI networks obtained by
different experimental methods for PPI detection.  Given a network, we aim to de-noise the
network, with the goal of determining which of all pairs of nodes in
the network should be connected by edges, with respect to a variety of
existing as well as new  LP measures.  We evaluate the different measures in
systematic precision-recall and receiver-operator curve frameworks. The details are as follows.

\subsection*{Network data}\label{sect:data_sets}

We evaluate all LP methods on three \emph{S.  cerevisiae} yeast PPI
networks obtained with different experimental methods.  We study PPI
networks of \emph{yeast} because yeast has been the most studied
species to date. As such, it has the most complete interactome and
thus represents the best species to evaluate the methods on.  We study
\emph{multiple} yeast PPI networks obtained with \emph{different}
experimental methods for PPI detection to test whether LP results are
dependent on the experimental method.  The three networks are: 1)
\emph{Y2H} network, obtained by Y2H, which consists of 1,647 nodes and
2,518 edges \cite{Yu2008,Solava2012}; 2) \emph{AP/MS} network,
obtained by AP/MS, which consists of 1,004 nodes and 8,319 edges
\cite{Yu2008,Solava2012}; and 3) \emph{high-confidence (HC)} network,
obtained from multiple data sources, which consists of 1,004 nodes and
8,323 edges \cite{Collins07}.  The quality of PPIs in the HC network
is comparable to the quality of interactions produced by precise
small-scale biological experiments \cite{Collins07}. Importantly, in
addition to the high-confidence PPIs, the data by \cite{Collins07}
also contains the corresponding lower-confidence PPI data, which is
useful for evaluation of the LP methods (as explained below).

\subsection*{Existing commonly used LP
  measures}\label{sect:existing}

\subsubsection*{Degree-based measure} 

According to preferential attachment \cite{Newman2001}, the higher
the degrees of two nodes, the more likely the nodes are to interact.
The \emph{degree product (DP)} measure scores the potential edge
between two nodes $v$ and $w$ as: DP$(v,w) = d(v) \times d(w)$, where
$d(v)$ is the degree of node $v$ \cite{Lichtenwalter2010}.

\subsubsection*{Common neighbors-based measures} \label{sect:common_neighbors}

A popular idea is that the more neighbors two nodes share, the more
likely the nodes are to interact. 

The \emph{shared neighbors (SN)} measure scores the potential edge
between nodes $v$ and $w$ as: SN$(v,w) = |N(v) \cap N(w)|$, where
$N(v)$ is the set of neighbors of  $v$ \cite{Newman2001}.  SN
simply counts the shared neighbors.

\emph{Jaccard coefficient (JC)} scores the potential edge between two
nodes $v$ and $w$ as: JC$(v,w) = \frac{|N(v) \cap N(w)|}{|N(v) \cup
N(w)|}$ \cite{Salton1984}. That is, it scores two nodes with respect
to the size of their shared neighborhood relative to the size of their
entire neighborhoods combined.  As such, it favors node pairs for
which a high percentage of all neighbors are shared.

The \emph{Adamic-Adar (AA)} measure scores the potential edge between
two nodes $v$ and $w$ as: AA$(v,w) = \sum_{z\in N(v) \cap N(w)}
\frac{1}{d(z)}$ \cite{Adamic2003}. Thus, of all common neighbors of 
two nodes, it favors low-degree shared neighbors over high-degree
shared neighbors.

\subsection*{New LP measures}\label{sect:new}

We already designed sensitive measures of topology that unlike the
existing measures go beyond capturing only the direct neighborhoods of
nodes to be linked. We used them for network alignment
\cite{GRAAL,HGRAAL,MIGRAAL}, clustering
\cite{Milenkovic2008,MMGP_Roy_Soc_09,Ho2010,Solava2012}, and
modeling \cite{Memisevic10a, Milenkovic2009}, but they
have not been used for LP thus far. Thus, we introduce them as new LP
measures.  Also, we design conceptually new
measures. The details are as follows.

\subsubsection*{Existing sensitive measures of topology as new LP measures}\label{sect:new_1}
 
To go beyond capturing only the direct network neighborhood of a node,
we previously designed a constraining graphlet-based measure of
topology, called \emph{node graphlet degree vector (node-GDV)}, that
captures up to 4-deep neighborhood of a node; a graphlet is a small
induced subgraph of the network \cite{Przulj04}. We designed a
measure of topological similarity of such extended neighborhoods of
two nodes, called \emph{node-GDV-similarity}. In this study, we use
node-GDV-similarity for LP, with the hypothesis that the more
topologically similar two nodes are, the more likely the nodes are to
interact.  Also, since shared neighbors-based approaches, which are
among the best LP measures over the widest range of real-world
networks
\cite{Lichtenwalter2010}, are based on the number of 3-node paths
that two nodes in question share, where a 3-node path is just a 3-node
graphlet, we generalize these measures by counting the number of all
3-5-node graphlets that the two nodes share.  We do this by using a
sensitive measure called \emph{edge-GDV}.  The formal description of
all of the measures is as follows.

\begin{figure}[!h]
\begin{center}
\includegraphics[width=4in]{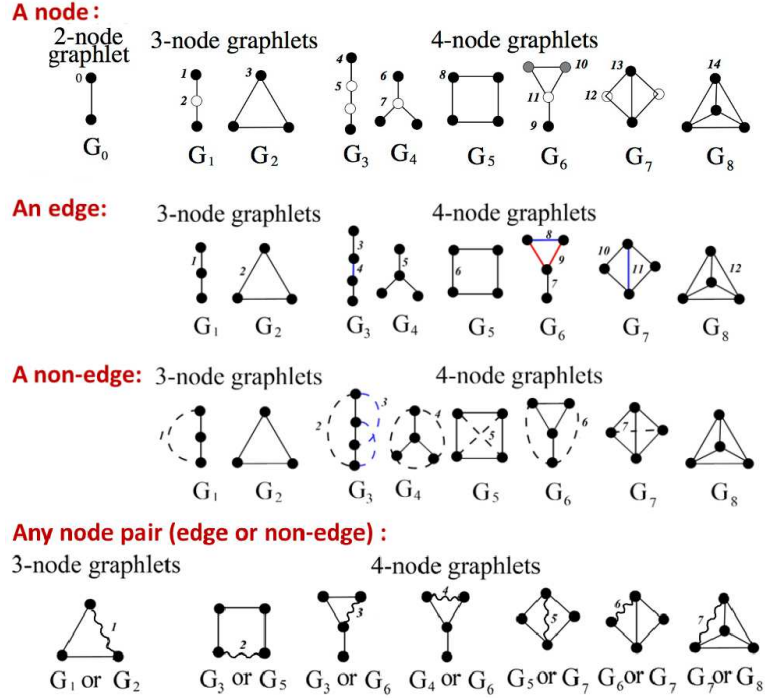}
\end{center}
\caption{
{\bf Graphlet positions of a node, an edge, a non-edge, and a node pair.}  All topological positions (``orbits'') in up to 4-node
  graphlets of a node (top; node shade), an edge (upper middle; solid
  line), a non-edge (lower middle; broken line), and any node pair, an
  edge or a non-edge (bottom; wavy line) are shown.  For example: 1) in graphlet
  $G_3$, the two end nodes are in node orbit 4, while the two middle
  nodes are in node orbit 5; 2) in $G_3$, the two ``outer'' edges are
  in edge orbit 3, while the ``middle'' edge is in edge orbit 4; 3) in
  $G_3$, the non-edge touching the end nodes is in non-edge orbit 2,
  while the two non-edges that touch the end nodes and the middle
  nodes are in non-edge orbit 3; 4) a node pair at node pair orbit 1
  touches a $G_2$ at edge orbit 2, if it is an edge, or a $G_1$ at
  non-edge orbit 1, if it is a non-edge (hence, mutually exclusive
  edge orbit 2 and non-edge orbit 1 are reconciled into a common node
  pair orbit 1). There are 15 node, 12 edge, 7 non-edge, and 7 node
  pair orbits for up to 4-node graphlets. In a graphlet, different
  orbits are colored differently.  All up to 5-node graphlets are
  used, but only up to 4-node graphlets are illustrated.  There are 73
  node, 68 edge, 49 non-edge, and 49 node pair orbits for up to 5-node
  graphlets.
}
\label{fig:graphlets_all}
\end{figure}

\vspace{0.1cm} \hspace{-0.55cm}\textbf{Node graphlet degree vector
  (node-GDV).}  We generalized the degree of node $v$ that counts the
number of edges that $v$ touches (where an edge is the only 2-node
graphlet, denoted by $G_0$ in Fig.  \ref{fig:graphlets_all}), into
\emph{node-GDV} of $v$ that counts the number of 2-5-node graphlets
that $v$ touches \cite{Milenkovic2008}.  We need to distinguish
between $v$ touching, for example, a three-node path ($G_1$ in Fig.
\ref{fig:graphlets_all}) at an end node or at the middle node, because
the end nodes are topologically identical to each other, while the
middle node is not.  This is because an automorphism (defined below)
of $G_1$ maps the end nodes to one another and the middle node to
itself.  Formally, an isomorphism $f$ from graph $X$ to graph $Y$ is a
bijection of nodes of $X$ to nodes of $Y$ such that $xy$ is an edge of
$X$ if and only if $f(x)f(y)$ is an edge of $Y$. An automorphism is an
isomorphism from X to itself. The automorphisms of $X$ form the
automorphism group, $\mbox{Aut}(X)$. If $x$ is a node of $X$, then the
automorphism node orbit of $x$ is $\mbox{Orb}_n(x) = \{ y \in V(X) | y =
f(x) \mbox{ for some } f \in \mbox{Aut}(X)\}$, where $V(X)$ is the set
of nodes of $X$.
There are 73 node orbits for 2-5-node graphlets.  Hence, node-GDV of
$v$ has 73 elements counting how many node orbits of each type touch
$v$ ($v$'s degree is the first element).  It captures $v$'s up to
4-deep neighborhood and thus a large portion of real networks, as they
are small-world \cite{Watts-Strogatz98}.

\vspace{0.1cm} \hspace{-0.55cm}\textbf{Node-GDV-similarity.}  To
compare node-GDVs of two nodes, one could use some existing measure,
e.g., Euclidean distance. However, this might be inappropriate, as
some orbit counts are not independent. Hence, we designed a new
measure, called node-GDV-similarity, as follows
\cite{Milenkovic2008}. For a node $u \in G$, $u_i$ is the $i^{th}$
element of its node-GDV.  The distance between the $i^{th}$ orbits of
nodes $u$ and $v$ is $D_i(u,v) = w_i \times \frac{|log(u_i + 1) -
  log(v_i + 1)|}{log(max\{u_i, v_i\} + 2)}$, where $w_{i}$ is the
weight of orbit $i$ that accounts for orbit dependencies
\cite{Milenkovic2008}. The $log$ is used because the $i^{th}$ elements
of two node-GDVs can differ by several orders of magnitude and we did
not want the distance between node-GDVs to be dominated by large
values.  The total distance is $D(u,v) =
\frac{\sum_{i=0}^{72}D_i}{\sum_{i=0}^{72}w_i}$. Finally,
node-GDV-similarity is $S(u,v) = 1 - D(u,v)$.  The higher the
node-GDV-similarity between nodes, the higher their topological
similarity.

\vspace{0.1cm} \hspace{-0.55cm}\textbf{Edge-GDV.}  
Since a graphlet contains both nodes \emph{and} edges, we defined
\emph{edge-GDV} to count the number of graphlets that an \emph{edge}
touches at a given ``edge orbit'' (Fig.  \ref{fig:graphlets_all})
\cite{Solava2012}.
Given the automorphism group of graph $X$, $\mbox{Aut}(X)$, if $xy$ is
an edge of $X$, the edge orbit of $xy$ is $\mbox{Orb}_e(xy) = \{ zw
\in E(X) | z = f(x) \mbox{ and } w = f(y) \mbox{ for some } f \in
\mbox{Aut}(X)\}$, where $E(X)$ is the set of edges of $X$.  There are  
68 edge orbits for 3-5-node graphlets \cite{Solava2012}.
(We designed \emph{edge-GDV-similarity} to measure topological
similarity of \emph{edges}, which we used for network clustering
\cite{Solava2012}.  However, we do not use this measure for LP.)

\subsubsection*{Conceptually novel measures of
  topology}\label{sect:new_2}

We need to predict the existence of a link between nodes independent
on whether there is an edge between them in the original network or
not.  Thus, in addition to describing the network position of an edge,
we need to be able to describe the position of a non-edge as well.
Hence, we generalize edge-GDV into \emph{non-edge-GDV} to measure the
topological position of a non-edge.  Then, we reconcile mutually
exclusive edge-GDVs and non-edge-GDVs into a new \emph{node-pair-GDV}
measure, which counts the number of graphlets that a node pair (an
edge or a non-edge) touches at a given ``node pair orbit'' (defined
below).  Finally, based on node-pair-edge-GDV of a node pair, we
create a new measure of the topological centrality of the node pair,
called \emph{node-pair-GDV-centrality}.  According to this measure,
the more graphlets the two nodes participate in (or share), the higher
their centrality.  Then, node-pair-GDV-centrality is used as a LP
measure to score potential edges between node pairs in the network.
The measures are defined as follows.

\vspace{0.1cm} \hspace{-0.55cm} \textbf{Non-edge-GDV.}  Analogous to
edge-GDV, in this study, we define \emph{non-edge-GDV} to count the
number of graphlets that a \emph{non-edge} touches at a given
``non-edge orbit'' (Fig. \ref{fig:graphlets_all}).  We define
non-edge orbits as follows.  If $xy$ is a non-edge of graph $X$, then
the non-edge orbit of $xy$ is $\mbox{Orb}_{ne}(xy) = \{ zw \in C(X) |
z = f(x) \mbox{ and } w = f(y) \mbox{ for some } f \in
\mbox{Aut}(X)\}$, where $C(X)$ is the set of all non-edges of $X$.
For example, in Fig. \ref{fig:graphlets_all}, in graphlet $G_1$,
the only non-edge is in non-edge orbit 1.  Graphlet $G_2$ has no
non-edges. In graphlet $G_3$, the non-edge that touches the two end
nodes is in one non-edge orbit (non-edge orbit 2), while the remaining
two non-edges that touch the end nodes and the middle nodes are in a
different non-edge orbit (non-edge orbit 3).  And so on.  There are 49
non-edge orbits for 3-5-node graphlets.

\vspace{0.1cm} \hspace{-0.55cm}\textbf{Node-pair-edge-GDV.}  Edge and
non-edge orbits are mutually exclusive (Fig.
\ref{fig:graphlets_all}).  However, to perform LP, we
need to contrast the topological neighborhood of nodes $v$ and $u$
against the neighborhood of nodes $s$ and $t$, while hiding the
information about whether $v$ and $u$ or $s$ and $t$ are actually
linked.  Hence, we need to reconcile edge orbits and non-edge orbits
by defining \emph{node-pair-GDV} to count the number of graphlets that
a general \emph{node pair}, which can be either an edge or a non-edge,
touches at a given ``node pair orbit''.  For example, in Fig.
\ref{fig:graphlets_all}, a node pair at node pair orbit 1 touches
a triangle (graphlet $G_2$) at edge orbit 2, if the node pair is an
edge, or it touches a three-node path (graphlet $G_1$) at non-edge
orbit 1, if the node pair is a non-edge.  Hence, we reconcile mutually
exclusive edge orbit 2 and non-edge orbit 1 into a common node pair
orbit 1.  We do this for all edge- and non-edge orbits, resulting in
49 node pair orbits for 3-5-node graphlets.

\vspace{0.1cm} \hspace{-0.55cm} \textbf{Node-pair-GDV-centrality.}  We
design \emph{node-pair-GDV-centrality} to assign high centrality
values to node pairs that participate in many graphlets. For nodes $v$
and $u$, if $c_i$ is the $i^{th}$ element of node-pair-GDV of the two
nodes, then $node$-$pair$-$GDV$-$centrality(vu) = \sum_{i=0}^{49} w_i
\times log(c_i + 1)$.  Thus, the more graphlets a node pair
participates in, the higher its centrality. Note that we previously
designed an analogous measure of the network centrality of a node,
called node-GDV-centrality \cite{Milenkovic2011}.

\subsubsection*{Using the new measures for LP}\label{sect:new_3}

Node-GDV-similarity and node-pair-GDV-centrality measures
allow for several simple modifications
which could perhaps improve LP accuracy, as follows.

\vspace{0.1cm} \hspace{-0.55cm}\textbf{Combining node-GDV-similarity
  and node-pair-GDV-centrality.}  Node-GDV-similarity favors linking
topologically similar nodes.
Node-pair-GDV-centrality favors linking nodes that share many
graphlets. Combining the two would favor linking nodes that are
\emph{both} topologically similar and share many graphlets. We combine
them
as: $(1-\alpha)\times node$-$GDV$-$similarity +
\alpha \times node$-$pair$-$GDV$-$centrality$. We vary $\alpha$
from $0$ to $1$ in increments of 0.2.

\vspace{0.1cm} \hspace{-0.55cm}\textbf{Prioritizing dense graphlets.}
Node-pair-GDV-centrality, as defined above, counts the number of
graphlets that two nodes share, while assigning weights to different
graphlets only with respect to ``orbit dependencies'' (see
\cite{Milenkovic2008} for details). However, it ignores any
information about the \emph{denseness} of the graphlets that the two
nodes share. Analogous to Adamic-Adar which favors some shared
neighbors over others based on their degrees (see above), we might want to favor some shared
graphlets over others based on their denseness.  For example, it might
be more reasonable to link two nodes that share many 4-node cliques
than two nodes that share many 4-node paths. So, we favor denser
shared graphlets over sparser shared graphlets by defining
\emph{density-weighted} (or simply \emph{weighted})
\emph{node-pair-GDV-centrality} (Supplementary Section S1).  We
evaluate both unweighted and weighted node-pair-GDV-centrality
measures.

\vspace{0.1cm} \hspace{-0.55cm}\textbf{Graphlet size.} To test how
much of network topology is beneficial for LP, when using the
graphlet-based measures, we use: 1) all 3-5-node graphlets, 2)
3-4-node graphlets, but not 5-node graphlets, and 3) only 3-node
graphlets. Note that using the only 3-node graphlet within the
node-pair-GDV-centrality at $\alpha$ of 1 (see above) is equivalent to
the SN measure (see above).  Hence, SN is a
variation of node-pair-GDV-centrality.  Also, note that when using
3-node graphlets, unweighted and weighted
node-pair-GDV-centralities are equivalent. This is because there is
only one 3-node graphlet when dealing with node-pair-GDVs, and its
density is one.  Determining which amount of topology to use is
important: the more topology (the larger the graphlets), the higher
the computational complexity.  Exhaustive counting of all graphlets on
up to $n$ nodes in graph $G(V,E)$ takes $O(|V|^n)$; but, the practical
running time is much smaller due to the sparseness of real networks
\cite{GraphCrunch,Przulj06ECCB,Marcus2011}.  Also, counting is
embarrassingly parallel.  Finally, fast non-exhaustive approaches
exist for counting graphlets \cite{Marcus2011}.

\subsection*{Evaluation framework}\label{sect:evaluation}

We evaluate each of the existing and new LP methods  on each of the PPI networks as follows.

First, we introduce synthetic noise in the given PPI network by
randomly removing 5\%-50\% of its edges, with the goal of measuring
how well the different methods can reconstruct the original network,
using the original PPIs as the ground truth data. We apply the given
LP measure to a ``noisy'' network created in this way and score each
node pair in the network, so that the higher the score, the more
likely the nodes are to be linked.  We predict $k\%$ of the
highest-scoring node pairs as edges.  We vary $k$ from 0\% to 100\% in
increments of 1\%.  At each $k$, we count the number of true
positives, true negatives, false positives, and false negatives,
and
we compute: 1) precision, recall, and F-score; and 2) sensitivity and
specificity (Supplementary Section S2) \cite{Davis06}.  For
simplicity of comparing results across different methods, we summarize
the performance of the methods over the entire range of $k$ with
respect to sensitivity and specificity by calculating the areas under
receiver-operator curves (AUROCs).

To account for randomness in the above procedure, for each level of
noise, we randomly remove the given percentage of edges from the
original network five times and average the above statistics over the
five runs. Ideally, we would perform more random runs, but this is
impractical due to the required computational time. Plus, this might
be unnecesary, since the standard deviations resulting from the five
runs are typically very small (Section ``Results and Discussion''), and since
even with five random runs of each method, we can compute the
statistical significance of the difference in LP accuracy between a
pair of methods by using the \emph{paired} $t$-test. With this test,
we compare five pairs of AUROCs corresponding to five random runs of
two methods, and a low $p$-value would indicate that the null
hypothesis (the difference between the acuracy of the two methods
having a mean of 0) can be rejected.

Second, due to the availability of low-confidence PPI data for the HC
network (see above), we perform an additional
evaluation test: we apply the given LP method to the HC network and
use the low-confidence PPIs as the ground truth data.  We evaluate the
method in the same way as above.

Third, we apply the given LP method to a network to de-noise it, and
we evaluate the biological quality of the de-noised network with
respect to the ``enrichment'' of predicted edges in Gene Ontology (GO)
terms \cite{Go00}. We compute the enrichment as the percentage of
predicted edges, out of all edges in which both proteins have at least
one GO term, in which the two end nodes share a GO term.  As
\cite{Kuchaiev2010}, we do this for biological process GO
terms. To avoid potential biases, we consider only gene-GO term
associations with experimental evidence codes.
Since we de-noise networks by relying on their topology (i.e., on the
PPIs), to avoid ``circular arguments'', of these associations, we
exclude associations inferred from PPIs.  We compute the statistical
significance of the enrichment
by using the hypergeometric model (Supplementary Section S2).

Finally, we validate predicted edges absent from the original network
by searching for them in an independent PPI data source.  Here, we use
BioGRID \cite{BIOGRID}, because it is a trusted PPI data source.
Again, we measure the statistical significance of validating the given
number of predictions by using the hypergeometric model (Supplementary
Section S2).
We perform the external data source validation on AP/MS predictions as
this network uses the same naming scheme as BioGRID.

\section*{Results and Discussion}\label{sect:results}

We study three yeast PPI networks: AP/MS, Y2H, and HC. We use a number of existing  and new LP measures.
The existing measures are degree product (DP), shared neighbors (SN),
Jaccard coefficient (JC), and Adamic-Adar (AA). The new measures are node-GDV-similarity
 and node-pair-GDV-centrality. See Methods for details.

The two graphlet-based measures allow us to address several important
LP questions. First, we can combine node-GDV-similarity, which favors
linking nodes with topologically similar neighborhoods, with
node-pair-GDV-centrality, which favors linking nodes that share many
graphlets and thus have large \emph{extended} shared neighborhoods, to
favor linking nodes that are \emph{both} topologically similar and
share many graphlets, which might be preferred.  To test whether this
is the case, we combine the two measures by varying the value of
parameter $\alpha$ from 0 to 1, where $\alpha$ of 0 means that only
node-GDV-similarity is used, and $\alpha$ of 1 means that only
node-pair-GDV-centrality is used (see Methods).  Second,
we test whether favoring denser graphlets that are shared between the
nodes in question within the node-pair-GDV-centrality measure is
preferred over equally favoring all graphlets, independent on their
density. We do this by evaluating both unweighted and weighted
node-pair-GDV-centrality measures (see Methods). Third, to test how much of network topology is
beneficial for LP, when using the graphlet-based measures, we use: 1)
all 3-5 node graphlets, 2) only 3-4-node graphlets, and 3) only 3-node
graphlets.

After we compare the different variations of graphlet-based measures,
we evaluate the best of them against the existing measures.  All methods are
evaluated fairly and systematically in AUROC and precision-recall
settings (see Methods).  We perform two evaluation
tests: 1) we introduce synthetic noise in the given PPI network by
randomly removing a percentage of its edges, with the goal of
measuring how well the given method can reconstruct the original
network, using the original PPIs as the ground truth data; and 2) given the availability of
low-confidence PPI data corresponding to the HC network, we apply the given LP method to the original HC
PPI network and use the corresponding low-confidence PPI data as the
ground truth data when evaluating the method (see Methods and below for details).

After we \emph{evaluate} the methods, we \emph{apply} them to the PPI
networks to de-noise them, and we evaluate biological quality of the
de-noised networks: 1) with respect to the ``enrichment'' of predicted
edges in GO terms \cite{Go00}, and 2) by validating predicted edges
in an external data source (see Methods and below for details).  
Ultimately, we are less focused on identifying a superior LP method
but more on testing whether we can de-noise a network so that the
de-noised network is biologically more meaningful than the original
one, as well as on which topological properties affect LP accuracy.

\subsection*{Evaluating LP methods by introducing synthetic noise into
  PPI networks}\label{sect:synthetic_networks}

Current PPI networks are noisy. The correct and complete ground truth
interactomes are unknown. Thus, an alternative ground truth data has
to be sought. We create synthetic ground truth data from the real PPI
networks.  For each PPI network, we add synthetic noise to the network
by randomly removing 5\%, 10\%, 15\%, 20\%, 25\%, and 50\% of the
original edges. Then, we evaluate the given LP method by applying it
to a synthetically noised network and by measuring how well it
reconstructs the original network (see Methods).

\subsubsection*{Combining topological similarity and centrality of nodes to be linked improves LP accuracy}\label{sect:similarity}

\begin{figure}[!h]
\begin{center}
\includegraphics[width=6in]{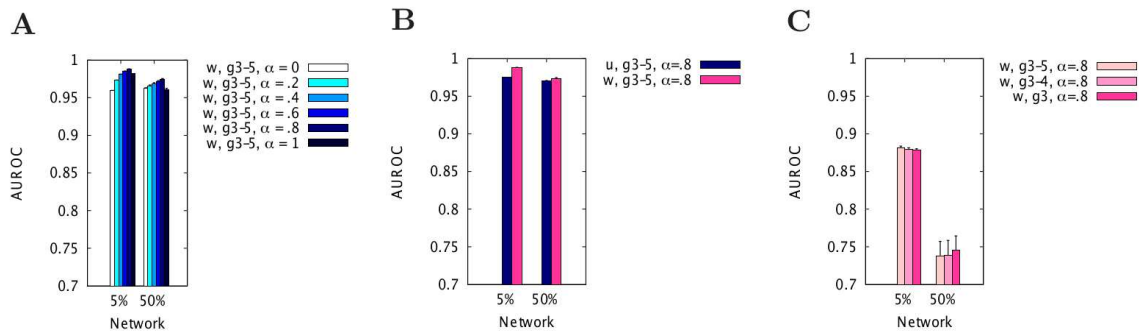}
\end{center}
\caption{
{\bf LP accuracy of the graphlet-based methods.}  The accuracy is shown in terms
of AUROCs  at the lowest noise level of 5\% and the highest noise level of 50\%
  when comparing: \textbf{A)} varying $\alpha$s in the AP/MS network;
  \textbf{B)} unweighted (``u'') vs.  weighted (``w'')
  node-pair-GDV-centrality in the HC network; and \textbf{C)}
  different graphlet sizes (3-5-node (``g3-5''), 3-4-node (``g3-4''),
  and 3-node (``g3'') graphlets) in the Y2H network.  Note that we
  intentionally vary the networks between panels (AP/MS in panel A,
  HC in panel B, and Y2H in panel C), but only in order to
  represent each of the three studied networks equally; we show full
  results in Supplementary Fig.  S1-S4. 
}
\label{fig:AUROC}
\end{figure}

By combining node-GDV-similarity and node-pair-GDV-centrality with
parameter $\alpha$ (see Methods), we find that nodes that
are simultaneously topologically similar and share many graphlets are
preferred for LP.  In general, the larger the value of $\alpha$ (the
more node-pair-GDV-centrality is used), the better the LP accuracy
(Fig. \ref{fig:AUROC} A).  This suggests that the topological
similarity of two nodes is less relevant for LP than the number of
graphlets that the nodes share.  However, using a small amount of
node-GDV-similarity in the combined LP score ($\alpha=0.8$) actually
improves LP accuracy compared to using node-pair-GDV-centrality alone
($\alpha=1$), implying that topological similarity \emph{is} relevant.
The difference between LP accuracy of the best $\alpha$ of $0.8$ and
any other $\alpha$ in Fig. \ref{fig:AUROC} A is statistically
significant, with $p$-values below $2.5
\times 10^{-6}$ for  5\% noise and below  $2.8
\times 10^{-4}$ for  50\% noise.

While the results in Fig.  \ref{fig:AUROC} A are for weighted
node-pair-GDV-centrality, 3-5-node graphlets, two noise levels, and
the AP/MS network, in general, they also hold for weighted
node-pair-GDV-centrality, all graphlet sizes, all noise levels, and HC
and Y2H networks (Supplementary Fig. S1 and S2). And since
$\alpha = 0.8$ is statistically significantly superior to all other
$\alpha$s, in the rest of the section, we focus only on this value of
$\alpha$.

\subsubsection*{Favoring denser shared graphlets improves LP
  accuracy}\label{sect:weighting}

We find that preferring denser graphlets (see Methods)
improves the LP performance: weighted node-pair-GDV-centrality
outperforms the unweighted version (Fig.  \ref{fig:AUROC} B), and
its superiority is statistically significant, with $p$-value of $1.76
\times 10^{-8}$ for 5\% noise and  $1.8
\times 10^{-5}$ for 50\% noise.

While the results in the figure are for all 3-5-node graphlets, two
noise levels, and the HC network, in general, they also hold for all
graphlet sizes, noise levels, and AP/MS and Y2H networks
(Supplementary Fig. S3). Thus, henceforth, we focus only on the
superior weighted version of node-pair-GDV-centrality.

\subsubsection*{Using more topology does not always guarantee higher LP
  accuracy}\label{sect:graphlet_size}

There is no clear trend on how much topology is best (Fig.
\ref{fig:AUROC} C). For example, for the lowest noise level of 5\%, using 3-5-node 
graphlets is statistically significantly superior over using 3-4-node
or 3-node graphlets, with $p$-values of $5.8 \times 10^{-3}$ and $1.1
\times 10^{-2}$, respectively. On the other hand, for the highest 
noise level of 50\%, using 3-5-node graphlets is marginally superior
over using 3-4-node graphlets, with $p$-value of $5.6\times 10^{-2}$,
and it is statistically significantly superior over using 3-5-node
graphlets, with $p$-value of $8.6 \times 10^{-3}$. Hence, using more
network topology can improve LP accuracy, but it is not guaranteed to
do so.

Whereas the results in Fig.  \ref{fig:AUROC} C are only for two
noise levels and the Y2H network, in general, they also hold for other
noise levels and for AP/MS and HC networks (Supplementary Fig.  S4).

Because in some cases using only 3-node graphlets is superior, and
because the existing shared neighbors-based methods and SN in
particular also rely on 3-node graphlets (see Methods),
one might incorrectly assume that in these cases, our graphlet-based
methods do not improve upon the existing methods. However, it is at
$\alpha$ of 1 when our node-pair-GDV-centrality and existing SN method
are equivalent (see Methods). Since our results at
$\alpha$ of 0.8 are superior over results at $\alpha$ of 1 (see above), and since at $\alpha$ of 0.8 SN is actually
combined with graphlet information encoded in the node-GDV-similarity
measure, node-GDV-similarity actually \emph{improves} the accuracy of
SN even when using 3-node graphlets only and especially when using all
3-5-node graphlets is superior to using only 3-node graphlets. 

Also, using deeper network topology is superior to using only the
direct network neighborhood of nodes to be linked in the sense that
node-GDV-similarity alone ($\alpha=0$) is superior to the existing DP
method (Supplementary Fig. S5). This is interesting because the two
methods are somewhat similar. They both take into account graphlet
degrees of two nodes in question. They differ in that DP considers
only the 2-node graphlet and hence captures only the direct (1-deep)
network neighborhoods of the nodes, whereas node-GDV-similarity
considers all 2-5-node graphlets, thus capturing up to 4-deep node
neighborhoods.  Hence, in this context, including more network
topology helps. (We
compare the existing methods to our new graphlet-based methods in more
detail in the following section.)

In general, using larger graphlets can increase LP accuracy (the following sections also confirm this).  Since counting larger graphlets is computationally expensive
compared to counting smaller graphlets (see Methods),
whether it is worth including the extra topological information that
is captured by the larger graphlets depends on how significant the improvement
is.

\subsubsection*{New graphlet-based measures are superior over existing measures}\label{sect:comparison}
  
Having examined different variations of the graphlet-based measures,
we now compare these measures with the existing ones. Of all
graphlet-based variations, we report the weighted version at $\alpha =
0.8$ when considering 3-5-node graphlets, since this version generally
performs the best.  The existing
measures include DP, SN, JC, and AA.

\begin{figure}[!h]
\begin{center}
\includegraphics[width=6in]{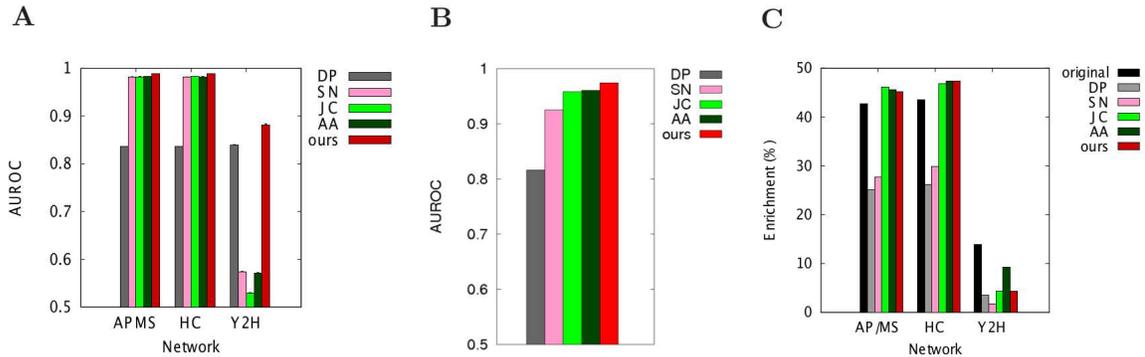}
\end{center}
\caption{
{\bf Comparison of different methods.} Our best method (``ours'') is compared against existing methods (DP, SN, JC, and AA) in
  terms of: \textbf{A)} AUROCs for synthetically noised AP/MS, HC,
  and Y2H networks at 5\% noise level; \textbf{B)} AUROCs for the HC
  network when using low confidence PPIs as the ground truth data; and
  \textbf{C)} GO enrichments of AP/MS, HC, and Y2H networks and their
  de-noised counterparts.  ``Ours'' corresponds to using 3-5-node
  weighted graphlets at $\alpha=0.8$ in panels A and C and using
  3-4-node weighted graphlets at $\alpha = 0.4$ in panel B.}
  \label{fig:compare}
\end{figure}

Our graphlet-based method is superior to all existing methods in all
three networks (Fig. \ref{fig:compare} A). The $p$-value of the
difference between LP accuracy of our method and any other method in
Fig. \ref{fig:compare} A is below $8.6 \times 10^{-6}$, $1.5 \times
10^{-6}$, and $1.2 \times 10^{-6}$ for AP/MS, HC, and Y2H networks,
respectively.  It is worth noting that most of the methods perform
quite well, reaching AUROCs of up to 0.99, 0.99, and 0.89 in AP/MS,
HC, and Y2H networks, respectively.  Whereas Fig. \ref{fig:compare}
A is for the lowest noise level, the results are similar for other
noise levels (Supplementary Fig. S5). Interestingly, our
graphlet-based measures further improve over the existing measures as
the noise increases.

\subsubsection*{AUROCs vs. precision-recall curves}\label{sect:ROC_PR}

Thus far, we have shown AUROC results.  AUROCs are commonly used to
evaluate methods over the entire [0\%,100\%] range of $k$
\cite{Lichtenwalter2010}, since the performance of each method can be
summarized into a single number over the entire range, while
precision-recall scores have to be shown for each value of $k$
(see Methods). However, we find that the AUROC
results are not necessarily consistent with precision-recall results.
While in both AP/MS and HC networks all weighted graphlet-based
measures are better than JC and AA with respect to AUROCs, JC and AA
are better with respect to precision-recall curves (Fig.
\ref{fig:compare} A and Supplementary Fig.  S5--S10).  Further, for
Y2H, we notice a different inconsistency: whereas AUROCs are high for
the best-performing methods (Supplementary Fig.  S5), precision-recall
curves indicate poor performance of all methods, as precision is
always low (Supplementary Fig. S8).

Even though optimizing AUROCs does not necessarily optimize
precision-recall \cite{Davis06}, the inconsistencies are alarming,
and the LP community needs to be aware.
We address this by also comparing the different methods with respect
to biological correctness of their de-noised networks (see below). But first, we check whether results
depend on the ground truth data, as follows.

\subsection*{Evaluating LP methods on HC network with respect to
  low-confidence PPI data}\label{sect:collins_2}

When we evaluate the LP methods on the HC network with low-confidence
PPIs as the ground truth data, we find that:

\begin{enumerate}

\item As in the previous section (when evaluating LP methods by introducing synthetic noise into PPI networks), combining topological
  similarity and centrality of nodes to be linked improves LP
  accuracy. However, now $\alpha$ of 0.4 is the best overall instead
  of $\alpha$ of 0.8 (Supplementary Fig. S15 and S16): topological
  similarity is now \emph{more} relevant than the number of shared
  graphlets.

\item As in the previous section, favoring denser graphlets
  improves LP accuracy for the best $\alpha$ (Supplementary Fig.
  S17).
  
\item As in the previous section, using more topology can
  improve LP accuracy (Supplementary Fig. S18). Using 3-4-node
  graphlets at the best $\alpha$ of 0.4 results in higher AUROC than
  using only 3-node graphlets at \emph{any} $\alpha$, and using
  3-5-node graphlets at $\alpha$ of 0.8 results in higher AUROCs than
  using only 3-node graphlets or 3-4-node graphlets at the same
  $\alpha$ (Supplementary Fig. S18).

\item As in the previous section, our best graphlet-based 
measure in this context (using 3-4-node weighted graphlets at $\alpha
= 0.4$) is superior to \emph{all} existing measures (Fig.
\ref{fig:compare} B).

\item We again see
 inconsistencies between AUROC and precision-recall results
 (Fig. \ref{fig:compare} B and Supplementary Fig. S19).

\end{enumerate}

\subsection*{De-noising the PPI
  networks}\label{sect:net-denoise-validate}

Since both our new and the existing methods perform well on all
networks with respect to AUROCs (Fig. \ref{fig:AUROC} A and B and
Supplementary Fig. S5), we use the overall best graphlet-based method
(weighted 3-5-node graphlets at $\alpha = 0.8$) as well as DP, SN, JC,
and AA to de-noise the networks.  We score each node pair in a
network, and we predict as edges in the de-noised network the top
$k\%$ highest scoring node pairs.  We choose $k$ so that the number of
edges in the de-noised network matches the number of edges in the
original network.  Depending on the network, $k$ falls between 1\% and
2\%.  We choose $k$ in this way because most of the methods achieve
the maximum F-score in this range of $k$ (Supplementary Fig.
S12-S14).

\subsubsection*{GO validation of de-noised
  networks}\label{sect:predictions}
We validate biological correctness of the de-noised networks by
computing the enrichment of all predicted edges in GO terms (see Methods
and Fig. \ref{fig:compare} C).  When we
de-noise AP/MS and HC networks, the enrichment is statistically
significant for all methods ($p$-values $\le 10^{-100}$).  Our method
as well as JC and AA \emph{improve} the quality of the original AP/MS
and HC networks. This is important, since the main goal of LP is to
de-noise a network so that the de-noised network is more meaningful
than the original one. While the GO enrichments are worse for the
de-noised networks than for the original Y2H network, the enrichments
are still statistically significant for our method, DP, JC, and AA
($p$-values $\le$ 0.05). Interestingly, in this context, some shared
neighbors-based measures slightly outperform our measure for AP/MS and
Y2H networks, but our measure is marginally better for the HC network
(Fig. \ref{fig:compare} C).

\subsubsection*{Intersection of de-noised networks produced by
  different methods}\label{sect:intersection}

Since we de-noise a network with multiple LP methods, we measure the
intersections between the de-noised networks (Supplementary Fig. S20).
The intersections are quite large between the shared neighbors-based
methods and our method.  JC is an exception, as it is somewhat
different not only from our measure but also from other shared
neighbors-based methods. Actually, our method is more similar to SN
and AA than JC is. But interestingly,
the intersections between the original networks and our method or JC
are slightly larger than the intersections between the original
networks and AA or SN, and all of these are much larger than the
intersections between the original networks and DP.

\subsubsection*{Validation of de-noised networks on external PPI
  data}\label{sect:validation}

We aim to validate ``new predicted edges'' (predicted edges not
present in the original network; Supplementary Table S1) by searching
for them in BioGRID as an independent data source. We do this for the
AP/MS network.
 Even though
validation accuracy varies across the methods, all methods achieve
statistically significant validation rates ($p$-values below $1\times
10^{-100}$).  Of the existing methods, only JC outperforms our
method  (Supplementary Fig. S21).

\section*{Concluding remarks}

We tackle the problem of link prediction (LP) in the context of PPI
network de-noising.  We comprehensively study what is it in the PPI
network topology around nodes in question that dictates whether the
nodes should be linked.  To evaluate whether nodes that share many
neighbors and are thus close in the network are favored over distant
nodes (as is the assumption of most of the existing LP methods),
whether topological similarity between nodes in question has any
effect, and how much of the network topology should be included, we
had to propose new LP methods, since none of the existing methods
allowed for answering all of these questions. Unlike the existing
methods, our new methods allow for combining topological similarity of
the nodes to be linked with the information about the size of their
shared neighborhood, and they allow for varying the amount of network
topology that is taken into account for LP.  After we demonstrate via
a thorough evaluation that our new methods are better than the
existing methods, we use the methods to de-noise PPI networks.
Importantly, the de-noised networks improve biological correctness of
the original networks, which is the ultimate goal of LP in
computational biology.

\section*{Acknowledgments}

We thank Nicholas J.  Taylor, an undergraduate Computer Science
student at the University of Notre Dame, for implementing the common
neighbors measures and running the corresponding analyses.

\bibliography{link_prediction_paper}

\end{document}